\documentclass[preprint,nofootinbib,12pt]{revtex4}

\usepackage{amsmath}
\usepackage{amsthm}
\usepackage{amsfonts}
\usepackage{graphicx}
\usepackage{psfrag}
\usepackage{verbatim}
\usepackage{tikz}

\usetikzlibrary{positioning}
\usetikzlibrary{decorations.pathmorphing}
\usetikzlibrary{decorations.markings}

\newcommand{\beq}{\begin{equation}}
\newcommand{\eeq}{\end{equation}}
\newcommand{\beqa}{\begin{eqnarray}}
\newcommand{\eeqa}{\end{eqnarray}}

\newcommand{\lsim}{\mathrel{\rlap{\lower4pt\hbox{\hskip1pt$\sim$}}
    \raise1pt\hbox{$<$}}}         
\newcommand{\gsim}{\mathrel{\rlap{\lower4pt\hbox{\hskip1pt$\sim$}}
    \raise1pt\hbox{$>$}}}         

\theoremstyle{definition} 
\theoremstyle{plain}

\begin{document}

\tikzset{ 
  photon/.style={decorate, decoration={snake,segment length=0.2cm}},
  electron/.style={postaction={decorate}, decoration={markings,mark=at
      position .5 with {\arrow{>}}}},
  gluon/.style={decorate, decoration={coil,amplitude=4pt, segment
      length=5pt}}
}


\vspace*{.0cm}

\title{Model of leptons from $SO(3) \to A_4$}

\author{Joshua Berger}\email{jb454@cornell.edu}
\author{Yuval Grossman}\email{yg73@cornell.edu}

\affiliation{\vspace*{4mm}Institute for High Energy Phenomenology\\
Newman Laboratory of Elementary Particle Physics\\ Cornell University,
Ithaca, NY 14853, USA\vspace*{6mm}}


\vspace{1cm}
\begin{abstract}
The lepton sector masses and mixing angles can be explained in models
based on $A_4$ symmetry. $A_4$ is a non-Abelian discrete group.
Therefore, one issue in constructing models based on it is explaining
the origin of $A_4$. A plausible mechanism is that $A_4$ is an
unbroken subgroup of a continuous group that is broken spontaneously.
We construct a model of leptons where the $A_4$ symmetry is obtained
by spontaneous symmetry breaking of $SO(3)$.
\end{abstract}

\maketitle

\section{Introduction}
Recent experiments
\cite{Apollonio:2002gd,Aharmim:2005gt,Ahn:2006zza,Abe:2006fu,:2008ee}
have given an increasingly accurate picture of the neutrino sector of
the new Standard Model ($\nu$SM). Current best measurements are
summarized in Table~\ref{data}.  These pieces of evidence paint a
picture radically different from that of the quark sector
\cite{Charles:2004jd,Bona:2005vz,Amsler:2008zzb} that exhibits
extremely small masses, small mass splittings and non-hierarchical
mixing angles.

Many attempts were made to obtain the masses and mixing angles from a
more fundamental theory. In this paper, we concentrate on the lepton
sector and consider the Pontecorvo-Maki-Nakagawa-Sakata (PMNS) mixing
matrix, $U$~\cite{Pontecorvo:1967fh,Maki:1962mu}.  Using the data
presented in Table \ref{data}, the current best fit for this matrix is
\begin{equation}
  |U| = 
  \begin{pmatrix}
    0.823 & 0.554 & 0.126\\
    0.480 & 0.558 & 0.677\\
    0.305 & 0.618 & 0.725
  \end{pmatrix}.
\end{equation}
It has been pointed out that within $2\sigma$ this matrix is
consistent with the Harrison-Perkins-Scott (HPS) mixing
matrix~\cite{Harrison:2002er}
\begin{equation}
  U_{\rm HPS} = \begin{pmatrix}
    \sqrt{\frac{2}{3}} & \frac{1}{\sqrt{3}} & 0\\
    -\frac{1}{\sqrt{6}} & \frac{1}{\sqrt{3}} & \frac{1}{\sqrt{2}}\\
    \frac{1}{\sqrt{6}} & -\frac{1}{\sqrt{3}} & \frac{1}{\sqrt{2}}\\
  \end{pmatrix}.
\end{equation}
The HPS matrix has a definite pattern.  This pattern has motivated
explanations of the structure of $U$ using non-Abelian discrete flavor
symmetries. Of particular interest is a class of models postulating an
$A_4$ family symmetry~\cite{Babu:2002dz,Altarelli:2005yp}.  In these
models, the left-handed lepton doublet and the right-handed neutrino
singlet transform in three dimensional irreducible representations,
while the right handed charged leptons transform under distinct one
dimensional representations.  For a review of such models, see
Ref. \cite{Altarelli:2009km}. Such models, however, are typically
plagued by two issues.

\begin{table}[!t]
  \begin{center}
    \begin{tabular}{|c|c l|}
      \hline
      ~Observable~ & Value & Main Source\\
      \hline
      $\sin^2 \theta_{12}$ & $0.312^{+0.019}_{0.018}$ & ~Solar neutrino
      experiments\\
      $\Delta m_{21}^2$ & $~7.67^{+0.16}_{-0.19} \times 10^{-5}~{\rm
eV}^2$ & ~\\
 \hline
      $\sin^2 \theta_{23}$ & $0.466^{+0073}_{-0.058}$ & ~Atmospheric neutrino
      experiments~\\
      $|\Delta m_{32}^2|$ & $~2.39^{+0.11}_{-0.08} \times 10^{-3}~{\rm
eV}^2$ & ~\\ \hline
      $\sin^2 \theta_{13}$ & $0.016^{+0.010}_{-0.006}$ & ~Global fit with
      all current data\\
      $\Delta m_{31}^2$ & $~2.39^{+0.11}_{-0.08} \times 10^{-3}~{\rm
        eV}^2$ & ~\\
      \hline
    \end{tabular}
    \caption{The current best fit values for neutrino mass splittings
      and mixing angles~\cite{Fogli:2008ig,Altarelli:2009wt}. All ranges are
      quoted at $1\sigma$.  }\label{data}
  \end{center}
\end{table}

The first is that of vacuum alignment.  The $A_4$ symmetry is broken
to $Z_3$ by a scalar $\phi$ that couples to the charged leptons and to
$Z_2$ by a scalar $\phi'$ that couples to the neutrinos. There is no
reason, \emph{a priori}, for this particular vacuum structure. One
approach to resolving this problem is to add scalars and symmetries,
possibly with supersymmetry, to enforce that vacuum
alignment~\cite{XSA4,Altarelli:2009kr}. Placing the scalars $\phi$ and
$\phi'$ on separate branes of an extra-dimensional
model~\cite{Altarelli:2005yp,Csaki:2008qq} is another possibility.

The second problem of $A_4$ based models, which is the problem that we
attempt to solve in this paper, is that of the origin of $A_4$.  The
symmetry group $A_4$ is chosen simply because it works, with no
motivation from UV physics.  This lack of motivation is exacerbated by
the fact that gravity is believed to break global symmetries.
Therefore, we look for possible motivations of the $A_4$ symmetry
group.  One possibility is that $A_4$ comes out as a subgroup of the
modular group~\cite{Altarelli:2005yx}, which often arises in string
theory.  Another possibility is that $A_4$ arises in the low-energy
effective theory obtained by orbifolds of a six dimensional
theory~\cite{Altarelli:2006kg}.  In this paper we present a model
where $A_4$ is obtained by spontaneously breaking a continuous
symmetry which we take to be the minimal choice, $SO(3)$. The idea of
embedding $A_4$ in $SO(3)$ has been discussed
in~\cite{Bazzocchi}. Unlike in our case, where the $SO(3)$ is
spontaneously broken, in~\cite{Bazzocchi} an explicit breaking of a
global $SO(3)$ symmetry was introduced.  The idea of spontaneously
breaking a continuous symmetry to a discrete subgroup has been
discussed in~\cite{Adulpravitchai:2009kd}. However, their procedure is
different then ours.

In section 2, we briefly review the general structure of models of
neutrino mixing using $A_4$ symmetry.  In section 3, we review the
vacuum structure of models with $SO(3)$ symmetry and how it can be
broken to $A_4$. In section 4 we construct a model
for the lepton sector based on spontaneously broken $SO(3) \to A_4$
symmetry. We conclude in section 5. Technical details are collected in
the appendices. In appendix A, we summarize important properties of
the group $A_4$ and introduce relevant group theory concepts.  In
appendix B, we describe one method for determining the vacua of a
theory with $SO(3)$ symmetry and a scalar transforming in the
$\mathbf{7}$ of the group.

\section{Models with $A_4$ symmetry}\label{a4models}

Implementing non-Abelian discrete flavor symmetries in a model
generically leads to patterns in the mass matrices.  These patterns
yield patterns in the mixing matrices after changing to the mass
basis.  It is natural to try to obtain $U_{\rm HPS}$ using such
symmetries.  In fact, several models \cite{TBA4,Altarelli:2009kr} did
it using $A_4$ symmetry.  These models have several common features
which we describe in this section.

We consider only the lepton sector. The basic required matter content
are the $\nu$SM fermions (including the RH singlet neutrinos), the SM
Higgs and two more scalars that are denoted by $\phi$ and
$\phi'$. The fermion field content is
\beq 
\psi_\ell(2,3)_{1/2}, \qquad \psi_e(1,1)_{-1}, \qquad
\psi_\mu(1,1')_{-1}, \qquad \psi_\tau(1,1'')_{-1},
\qquad \psi_n(1,3)_{0}, 
\eeq 
and the scalars are 
\beq 
H(2,1)_{1/2}, \qquad \phi(1,3)_0, \qquad \phi'(1,3)_0.  
\eeq 
We use standard notation, $(S,A)_Y$, where $S$ [$A$] is the
representation under $SU(2)_L$ [$A_4$] and $Y$ is the hypercharge. In
specific models more fields are added in order to satisfy vacuum
alignment conditions.  In addition, further symmetries are usually
required to forbid unwanted terms in the Lagrangian, as well as to
obtain the correct vacuum alignment.  The purpose of the two scalars
$\phi$ and $\phi'$ is to break the $A_4$ symmetry down to its $Z_3$
and $Z_2$ subgroups respectively.  For the standard basis described in
Appendix~\ref{app-math}, this breaking is achieved by the VEVs:
\begin{equation}
\langle\phi\rangle = (v,v,v),\qquad \langle\phi'\rangle =
(v',0,0).
\end{equation}
The two scalars are then made (by symmetries, for example) to couple
to different sectors of the model.  The $\phi$ couples to the charged
leptons, giving a $Z_3$ symmetric mass matrix, while the $\phi'$
couples to the neutrinos, giving a $Z_2$ symmetric mass matrix.

The Lagrangian for the fermions with the properties and fields
described above is:
\begin{equation}\label{model-lagr}
  \mathcal{L} = - \frac{y_e}{\Lambda} \overline{\psi}_\ell \phi H \psi_E
  - \frac{y_\mu}{\Lambda} (\overline{\psi}_\ell \phi)' H \psi_\mu -
  \frac{y_\tau}{\Lambda} (\overline{\psi}_\ell \phi)''H \psi_\tau -
  M\overline{\psi^c}_n \psi_n  - x_\nu \overline{\psi^c}_n
  \psi_n \phi' - y_\nu \overline{\psi}_\ell H \psi_n,
\end{equation}
where $(\overline{\psi}_\ell \phi)'$ [$(\overline{\psi}_\ell \phi)''$]
denotes that the product is taken such that the result transforms in
the $1'$ [$1''$]. This Lagrangian provides an effective description up
until a cutoff $\Lambda$. We assume that $M$ is much larger than the
weak scale. Notice that charged lepton masses would not be allowed
without including non-renormalizable operators. We did not include
terms that are suppressed by $1/\Lambda^2$.

We emphasize that the Lagrangian (\ref{model-lagr}) is not the most
general one.  It is missing several terms allowed by the symmetries
listed so far.  Any of the terms coupling to $\phi$ is allowed with
$\phi \to \phi'$ and vice-versa.  For example, $\overline{\psi^c}_n
\psi_n \phi$ is allowed.  This issue is generally solved by including
additional discrete or continuous Abelian symmetries.  For example,
ref. \cite{Altarelli:2009kr} describes a supersymmetric model with an
additional $Z_4$ and $U(1)_R$ symmetry under which $\phi$ and $\phi'$
transform differently.

The heavy neutrino states present due to the see-saw mechanism can be
integrated out.  The resulting low-energy Majorana mass matrix for the
neutrinos has the form
\begin{equation}\label{majneut-mass}
  m_\nu = \begin{pmatrix}
    a & 0 & 0\\
    0 & b & d\\
    0 & d & b\\
  \end{pmatrix},
\end{equation}
where $a$, $b$, and $d$ depend on the specifics of the model.  The
off-diagonal $d$ entries are a reflection of the $A_4 \to Z_2$
breaking. It is made possible by the fact that a singlet can be formed
out of the product of three triplets.  The mass matrix for the charged
leptons has the form
\begin{equation}\label{clepton-mass}
m_\ell = \begin{pmatrix}
    y_e & y_\mu & y_\tau\\ 
    y_e & y_\mu \omega & y_\tau \omega^2\\ 
    y_e & y_\mu\omega^2 & y_\tau\omega
\end{pmatrix},
\end{equation}
where $\omega\equiv e^{2\pi i/3}$ (see Appendix \ref{app-math} for more
details). This mass matrix is diagonalized by multiplying on the left
by
\begin{equation}\label{rotmat}
V = \frac{1}{\sqrt{3}}
 \begin{pmatrix}
  1 & 1 & 1\\
  1 & \omega & \omega^2\\
  1 & \omega^2 & \omega\\
\end{pmatrix}.
\end{equation}
The rotation matrix $V$ in (\ref{rotmat}) does not depend on any of
the parameters of the theory.  This fact helps ensure that no
hierarchy will appear in the neutrino mixing matrix.  No change of
basis is required for the right-handed leptons. Performing the full
diagonalization procedure, the physical PMNS matrix, $U$, is then
given by $U_{\rm HPS}$.

Specific implementations of the ideas described above have several
obstacles to overcome. First, in general, $A_4$ based models only
explain the mixing parameters and not the mass hierarchies. (Both
mixing and masses can be obtained in an RS-type
model~\cite{Csaki:2008qq}.)  Another issue, as we already discussed,
is the fact that extra symmetries are needed in order to forbid
problematic terms. There is also an issue of vacuum alignment, which
has been discussed in the introduction. Finally, there is the issue of
the origin for the $A_4$ symmetry group, which is the issue we discuss
in this paper.

\section{Spontaneous breaking of $SO(3)\to A_4$}\label{spont-break}

In order to motivate the use of $A_4$, we use a model where the group
$A_4$ arises from spontaneous breaking of a continuous symmetry. The
simplest choice of gauge group is
$SO(3)$~\cite{Etesi:1997jv,Ovrut:1977cn}.  We discuss the
representation necessary for a scalar to break $SO(3)$ to $A_4$ and
write down a potential for this scalar to demonstrate how spontaneous
symmetry breaking (SSB) is achieved.

\begin{table}[!t]
  \begin{center}
    \begin{tabular}{|c| c|}
      \hline
      ~Representation~ & ~Decomposition~\\
      \hline
      $\mathbf{1}$ & $\mathbf{1}$\\
      $\mathbf{3}$ & $\mathbf{3}$\\
      $\mathbf{5}$ & $\mathbf{3} + \mathbf{1'} + \mathbf{1''}$\\
      $\mathbf{7}$ & $\mathbf{3} + \mathbf{3} + \mathbf{1}$\\
      \hline
    \end{tabular}
    \caption{Decomposition of the four smallest representations of
      $SO(3)$ into irreducible representations of
      $A_4$.}\label{decomp}
  \end{center}
\end{table}

Let $T$ be a scalar that transforms under an irreducible
representation of $SO(3)$.  This irreducible representation of $SO(3)$
induces a representation of $A_4$ since $A_4$ is a subgroup of
$SO(3)$.  In Appendix \ref{app-math}, we write down a general method
for decomposing an irreducible representation of $SO(3)$ into
irreducible representations of $A_4$. The decomposition of the four
smallest representations is given in Table~\ref{decomp}.  For now, it
is important to note that the smallest non-trivial representation of
$SO(3)$ that contains a singlet of $A_4$ is the $\mathbf{7}$.  This is
the smallest representation that could in principle result in an $A_4$
invariant vacuum.  Thus, it is natural to start our attempt to
construct a model using a scalar in the $\mathbf{7}$.

A model with a scalar transforming in the $\mathbf{7}$ of $SO(3)$ has
been described in \cite{Ovrut:1977cn,Etesi:1997jv}.  We summarize the
results of \cite{Ovrut:1977cn}.  The $\mathbf{7}$ of $SO(3)$ can be
described by symmetric, traceless rank 3 tensors in 3D, denoted as
$T^{abc}$.  The most general renormalizable potential that can be
written is
\begin{equation}\label{pot-7}
V = -\frac{\mu^2}{2} T^{abc} T^{abc} + \frac{\lambda}{4} (T^{abc}
T^{abc})^2 + c\, T^{abc} T^{bcd} T^{def} T^{efa}.
\end{equation} 
Naively, there are other quartic terms that can be written down, but
they are linear combinations of the two quartic terms in
(\ref{pot-7}). Also note that cubic terms vanish since the cubic
singlet is formed by an antisymmetric product of identical fields. A
technique for minimizing the potential is presented in
Appendix~\ref{app-min}.  The results of the minimization are as
follows. In order to have a stable potential we need $\lambda>0$. In
order to have a VEV at all we require $\mu^2>0$. Then, the residual
symmetry depends on the relation between $c$ and $\lambda$. For
$c<-\lambda/2$, the potential becomes unstable. For $c>0$, the
residual symmetry is $D_3$. For $-\lambda/2 < c < 0$, the residual
symmetry is $A_4$. We learn that there is a large area in parameter
space where $SO(3)$ is broken to $A_4$. In our model we choose the
parameters such that this is the case.

\section{Model of lepton based on $SO(3)\to A_4$}\label{sec-model}

We move to describe the model. The symmetry of the model is 
\beq
SU(2)_L \times U(1)_Y \times SO(3)_F \times Z_2.
\eeq
At this stage we do not care if the $SO(3)_F$ is gauged or not.  For
the fermions, we consider only the leptons. The full matter content of
the scalar and lepton sectors of the model are summarized in
Table~\ref{fullmatt}. We also describe the symmetry breaking induced
by each of the scalars.

\begin{table}[!t]
  \begin{center}
    \begin{tabular}{|c|c|c|c|c|}
      \hline
      Field & $SU(2)_L$ & $U(1)_Y$ & $SO(3)_F$ & $Z_2$ \\
      \hline
      $\psi_\ell$ & $\mathbf{2}$ & $-1/2$ & $\mathbf{3}$ & $-$\\
      $\psi_f$ & $\mathbf{1}$ & $-1$ & $\mathbf{3}$ & $-$\\
      \hline
      $\psi_e$ & $\mathbf{1}$ & $-1$ & $\mathbf{1}$ & $+$\\
      $\psi_m$ & $\mathbf{1}$ & $-1$ & $\mathbf{5}$ & $+$\\
      $\psi_n$ & 1 & $0$ & $\mathbf{3}$ & $-$\\
      \hline
      $H$ & $\mathbf{2}$ & $1/2$ & $\mathbf{1}$ & $+$\\
      $\phi$ & $\mathbf{1}$ & $0$ & $\mathbf{3}$ & $-$\\
      $\phi'$ & $\mathbf{1}$ & $0$ & $\mathbf{3}$ & $+$\\
      $\phi_5$ & $\mathbf{1}$ & $0$ & $\mathbf{5}$ & $-$\\
      $T$ & $\mathbf{1}$ & $0$ & $\mathbf{7}$ & $-$\\
      \hline
    \end{tabular}\hspace*{35pt}
    \begin{tabular}{|c|c|c|}
      \hline Field & VEV & Invariant Subgroup \\ \hline 
      $H$ & $v_H$ & none \\
      $\phi$ & $(v,v,v)$ & $Z_3$ \\ 
      $\phi'$ & $(0,0,v')$ & $Z_2$ \\
      $\phi_5$ & $\begin{pmatrix} 0 & v_5 & v_5 \\ v_5 & 0 & v_5 \\
        v_5 & v_5 & 0 \\ \end{pmatrix}$ & $Z_3$ \\ 
      $T$ & $\sim v_T$ (see text) & $A_4$ \\ 

      \hline \end{tabular}
      \caption{Left: Matter content for the lepton and scalar sectors
        of the model.  The blocks contain the left-handed fermions,
        right-handed fermions, and scalars respectively.  Right:
        Vacuum expectation values for the scalars and the subgroup of
        $SO(3)_F$ under which they are invariant.  The $H$ gets the
        usual SM-like VEV and the $T$ gets a VEV as described in
        Section \ref{spont-break}}\label{fullmatt}
\end{center}
\end{table}

We start with the scalar sector of the model. There are five scalar
fields in the model. Three of them, $H$, $\phi$ and $\phi'$ are needed
in the $A_4$ model. When extending the model to an $SO(3)_F$ symmetry,
we add two scalars, $T$ and $\phi_5$. We need $T$ as it is responsible
for the $SO(3)_F \to A_4$ breaking. As we discuss later, $\phi_5$ is
needed because without it the tau and the muon would be degenerate.
In term of scales, things are simpler if we decouple the $SO(3)_F \to
A_4$ breaking (triggered by $v_T$) and the $A_4$ breaking (which is
done by $v$, $v'$ and $v_5$). That is, we assume the following
hierarchies of scales
\beq\label{scale-hier} 
\Lambda \gg v_T \gg v\sim v' \sim v_5 \gg v_H.
\eeq 
We do not try to explain these hierarchies.

Next, we discuss the fermions. The fields $\psi_\ell$, $\psi_e$, and
$\psi_n$ have the same representations under $SO(3)_F$ as under
$A_4$. They correspond directly to fields in the $A_4$
model. Complications arise when considering the right handed muon and
tau fields that transform as $\mathbf{1'}$ and $\mathbf{1''}$
respectively. The issue is that the $\mathbf{1'}$ and $\mathbf{1''}$
do not correspond to irreducible representations of $SO(3)_F$. Thus,
they must be obtained as parts of $SO(3)_F$ representations that
include extra singlets or triplets of $A_4$.  Further complications
arise from the fact that irreducible representations of $SO(3)$ are
real and, therefore, $\mathbf{1'}$ and $\mathbf{1''}$ must be part of
the same $SO(3)$ representation in the scenario with minimal matter
content.  The simplest choice of representation that contains both
$\mathbf{1'}$ and $\mathbf{1''}$ is the $\mathbf{5}$. This explains
why we introduce $\psi_m$, which is the field that after $SO(3)_F$
breaking gives us the right handed muon and tau fields.

A fermion that transforms in the $\mathbf{5}$ of $SO(3)_F$ decomposes
into pieces that transform under the $\mathbf{1'}$,
$\mathbf{1''}$, and $\mathbf{3}$ representations of
$A_4$.  A field transforming in the $\mathbf{5}$ can be written as a
traceless, symmetric matrix.  In this form, the decomposition is
\begin{equation}
\psi_m = \begin{pmatrix}
\psi_\mu + \psi_\tau & \psi_h^3 & \psi_h^2\\
\psi_h^3 & \omega \psi_\mu + \omega^2 \psi_\tau & \psi_h^1\\
\psi_h^2 & \psi_h^1 & \omega^2 \psi_\mu + \omega \psi_\tau\\
\end{pmatrix},
\end{equation}
where $\psi_\mu$ transforms as a $\mathbf{1'}$, $\mathbf{\psi_\tau}$
transforms as a $\mathbf{1''}$, and $\psi_h$ transforms as a
$\mathbf{3}$.  The use of a fermion in the $\mathbf{5}$ implies that
further matter content is required. Without it, we end up with extra
right-handed fields. These extra field can be ``removed'' by adding a
triplet left-handed fermion giving them a large Dirac mass. This is
the reason we add the left-handed triplet, $\psi_f$.

The most general Lagrangian, including $1/\Lambda$ terms, that is
responsible for charged lepton masses is given by
\beqa\label{fulllag} 
\mathcal{L} &=& - y_e \overline{\psi_\ell^a}
\frac{H}{\Lambda} \phi^a \psi_e - y_m \overline{\psi_\ell^a}
\frac{H}{\Lambda} \phi^b \psi_m^{ab} - y_m^T \overline{\psi_\ell^a}
\frac{H}{\Lambda} T^{abc} \psi_m^{bc} - y_e' \overline{\psi_f^a}
\phi^a \psi_e \nonumber \\ && - y_m' \overline{\psi_f^a} \phi^b
\psi_m^{ab} - y_m^{T\prime} \overline{\psi_f^a} T^{abc} \psi_m^{bc}
-y_m^5 \epsilon^{abc} \overline{\psi_\ell^a} \frac{H}{\Lambda}
\phi_5^{bd} \psi_m^{cd} -y_m^{5\prime} \epsilon^{abc}
\overline{\psi_f^a} \phi_5^{bd} \psi_m^{cd}.  
\eeqa 
The scalars get VEVs as indicated in Table~\ref{fullmatt}.  Consider
the masses of the charged fermions. There are six left-handed and six
right-handed fields that can mix. Working in the basis where the right
handed fields are
$(\psi_e,\psi_\mu,\psi_\tau,\psi_h^1,\psi_h^2,\psi_h^3)$ and the
left-handed ones are $(\psi_\ell,\psi_f)$ the mass matrix is roughly
\begin{equation}
m_\ell \sim \begin{pmatrix}
 v_H v/\Lambda & v_H v_T/\Lambda\\
  v & v_T
\end{pmatrix},
\end{equation}
where each block describes a $3 \times 3$ matrix. We see that there
are three heavy states (of order $v_T$), three light states (of order
$v_H v/\Lambda$), and that there is very small mixing between these
two sets of states. We identify the light states as the three charged
leptons, and we neglect the mixing between them and the heavy states.
This procedure leaves a charged lepton Dirac mass matrix of the form
(\ref{clepton-mass}), which is given by
\begin{equation}
m_\ell = \begin{pmatrix} 
y_e \frac{v_H v}{\Lambda} & y_m \frac{v_H v}{\Lambda} + y_m^5 (\omega^2 - \omega) \frac{v_H v_5}{\Lambda} &
  y_m \frac{v_H v}{\Lambda} + y_m^5 (\omega - \omega^2) \frac{v_H
    v_5}{\Lambda}\\ 
y_e \frac{v_H v}{\Lambda} & \omega[y_m \frac{v_H v}{\Lambda} + y_m^5 (\omega^2 - \omega) \frac{v_H v_5}{\Lambda}] &
  \omega^2 [y_m \frac{v_H v}{\Lambda} + y_m^5 (\omega - \omega^2) \frac{v_H
    v_5}{\Lambda}]\\ 
y_e \frac{v_H v}{\Lambda} & \omega^2[y_m \frac{v_H v}{\Lambda} + y_m^5 (\omega^2 - \omega) \frac{v_H v_5}{\Lambda}] &
  \omega [y_m \frac{v_H v}{\Lambda} + y_m^5 (\omega - \omega^2) \frac{v_H
    v_5}{\Lambda}]\\ 
\end{pmatrix},
\end{equation}
In order to diagonalize this matrix, we multiply on the left by $V$
introduced in (\ref{rotmat}). The resulting diagonal mass matrix for
the charged leptons is
\begin{equation}\label{mass-mat-diag}
m_\ell^{\rm diag}=
\begin{pmatrix}
\left|y_e \frac{v_H v}{\Lambda}\right| & 0 & 0\\ 
0 & \left|y_m \frac{v_H v}{\Lambda}
- y_m^5 i\sqrt{3} \frac{v_H v_5}{\Lambda}\right| & 0\\ 
0 & 0 & \left|y_m \frac{v_H v}{\Lambda} + y_m^5 i\sqrt{3} \frac{v_H
  v_5}{\Lambda}\right|
\end{pmatrix}.
\end{equation} 

Two remarks are in order regarding the mass matrix for the charged
leptons. First, note that the charged lepton scale is smaller then the
electroweak scale by a factor of $v/\Lambda$. This implies that $v$
and $\Lambda$ are at most a factor of $10^2$ apart. Recalling that we
assume that $\Lambda \gg v_T \gg v$, we conclude that the different
scales cannot be widely separated. That is, the ratio of scales is of
order ten. Since this ratio is not very large, the fact that we
neglected $1/\Lambda^2$ terms may not be justified.

The second remark is about the muon and tau masses. The matrix
(\ref{mass-mat-diag}) leads degenerate muon and tau if the
parameters of the theory are real or if $v_5=0$. Moreover, in order to
reproduce the observed ratio of masses, $m_\mu /m_\tau \sim 1/16$, some
amount of fine tuning is needed. Defining
\beq
a\equiv y_m v, \qquad b\equiv i \sqrt{3} y_m^5 v_5, \qquad
\alpha \equiv \arg(a b^*),
\eeq
we require
\beq
\frac{|a|^2+|b|^2-2 |a| |b| \cos\alpha}
{|a|^2+|b|^2+2 |a| |b| \cos\alpha} \sim \frac{1}{16^2}.
\eeq
That is, the phase between $y_m$ and $y_m^5$ must be very close to $\pi/2$
and the values of $a$ and $b$ must be very close to each other.  Given
this fine-tuning, it is clear that this model does not try to explain
the fermion mass hierarchy: the tuning of the scales of the charged
lepton sector is exchanged for a tuning of the scales $a$ and $b$ to
be very close to each other.

The neutrino sector works just as in the low-energy $A_4$ model
described in section \ref{a4models}.  Since the neutrinos are in
triplet representation, the Lagrangian is almost the same as in
Eq. (\ref{model-lagr}).  One issue is that the off-diagonal terms in
the Majorana mass matrix require a coupling to $T$. Coupling to $\phi$
and $\phi_5$ are forbidden by the $Z_2$ symmetry used to forbid terms involving
$\phi'$ in (\ref{fulllag}). Then the terms relevant for neutrino
masses are
\begin{equation}
\mathcal{L} = - M\overline{\psi^c}_n^a \psi_n^a - \frac{x_\nu}{\Lambda}
\overline{\psi^c}_n^a \psi_n^b \phi^{\prime c} T^{abc} 
- y_\nu \overline{\psi}_\ell^a H \psi_n^a.
\end{equation}
Recalling that $\phi'$ gets a VEV $(v',0,0)^T$ and $T^{abc}$ gets a
VEV $v_T x^{(a} y^b z^{c)}$, the neutrino Majorana mass matrix is
given by
\begin{equation}\label{numajmass}
m_\nu^M = \begin{pmatrix}
  M & 0 & 0 \\
  0 & M & x_\nu v' \frac{v_T}{\Lambda} \\ 
  0 & x_\nu v' \frac{v_T}{\Lambda} & M \\
\end{pmatrix},
\end{equation}
while the Dirac mass matrix is given by
\begin{equation}\label{nudirmass}
m_\nu^D = y_\nu v_H \begin{pmatrix}
  1 & 0 & 0\\
  0 & 1 & 0\\
  0 & 0 & 1\\
\end{pmatrix}.
\end{equation}
The low-energy effective Majorana matrix is then
\begin{equation}\label{numajmass-eff}
\tilde{m}_\nu^M = - m_\nu^D
(m_\nu^{M})^{-1} (m_\nu^D)^{-1}=\begin{pmatrix} 
  - y_\nu^2 \frac{v^2}{M} & 0 & 0 \\ 0
  & y_\nu^2 \frac{M v_H^2}{x_\nu v^{\prime2} v_T^2 - M^2 \Lambda^2} &
  y_\nu^2 x_\nu \frac{v_H^2 v' v_T}{M^2 \Lambda^2 - x_\nu^2
    v^{\prime 2} v_T^2} \\ 0 & y_\nu^2 x_\nu \frac{v_H^2 v'
    v_T}{M^2\Lambda^2 - x_\nu^2 v^{\prime 2} v_T^2} & y_\nu^2
  \frac{Mv_H^2}{x_\nu v^{\prime2} v_T^2 - M^2 \Lambda^2} \\
\end{pmatrix}.
\end{equation}
The matrix (\ref{numajmass-eff}) has precisely the form
(\ref{majneut-mass}).  Taking into account the action of $V$ on the
left-handed handed fields, it can then be diagonalized by rotating the
left-handed neutrinos by $U_{\rm HPS}$.  The resulting diagonal mass
matrix is
\begin{equation}\label{diag-nu-mass}
\tilde{m}_\nu^{\rm diag} = y_\nu^2 v_H^2 \begin{pmatrix}
  \frac{\Lambda}{M \Lambda + x_\nu v' v_T} & 0 & 0 \\
  0 & \frac{1}{M} & 0 \\
  0 & 0 & \frac{\Lambda}{M \Lambda - x_\nu v' v_T} \\
\end{pmatrix}.
\end{equation}

Two remarks are in order. First, we emphasis that the the result of
the diagonalization is that the physical PMNS matrix is given by the
HPS matrix, that is, $U=U_{\rm HPS}$. The second remark is about the
mass splittings in the neutrino sector.  The form of the neutrino
masses in (\ref{diag-nu-mass}) constraints the scales in the theory.
If $x_\nu v' v_T \ll M \Lambda$, then the splittings become very
small, in contradiction to the $\mathcal{O}(100)$ factor difference in
the measured values of $\Delta m_{12}^2$ and $\Delta m_{23}^2$. We
then conclude that $x_\nu v' v_T \sim M \Lambda$.  Since we require
$\Lambda \gg v_T \gg v'$ and perturbative Yukawa couplings, we
conclude that $v' \gg M$. This is not a problem, as both $v'$ and $M$
can be much above the weak scale.

\section{Discussion and Conclusions}

We have shown that, in principle, a model of lepton masses and mixings
using an $A_4$ discrete symmetry can be obtained by spontaneously
breaking a continuous symmetry. The model, however, is not very
elegant. We already mentioned the problem of the fine tuning required
to get the correct muon and tau masses. We discuss a few other
problems below.

The first issue is that of vacuum alignment in the full scalar
potential. In previous incarnations of the $A_4$ model, additional
symmetries and, often, scalars are needed in order to ensure the correct
vacuum alignment.  The question is even trickier in our case.  All
four scalars in the model need very specific alignments.  Without
additional symmetries, there are many couplings in the potential
between these scalars which affect the vacuum structure.  In
particular, a possibility is the case where the additional scalars
force the scalar $T$ away from the $A_4$ invariant vacuum.  With the
many additional degrees of freedom in this model, it is difficult to
verify the vacuum alignment or to correct the alignment if it does not
follow from the current iteration of the model.

\begin{figure}[!t]
  \begin{center}
    \begin{tikzpicture}
      \coordinate [label=135:$U(1)_Y$] (v1) at (2.5,0);
      \coordinate [label=90:$SO(3)_F$] (v2) at (4.5,1.5);
      \coordinate [label=-90:$SO(3)_F$] (v3) at (4.5,-1.5);
      \draw [photon] (0,0) -- (v1.center);
      \draw [photon] (v2.center) -- +(2.5,0);
      \draw [photon] (v3.center) -- +(2.5,0);
      \draw (v1.center) -- (v2.center) -- (v3.center) -- (v1.center);
    \end{tikzpicture}
    \caption{Triangle diagram contributing to a $U(1)_Y$ anomaly if
      the $SO(3)_F$ flavor symmetry is gauged.}\label{tri-anomaly}
  \end{center}
\end{figure}
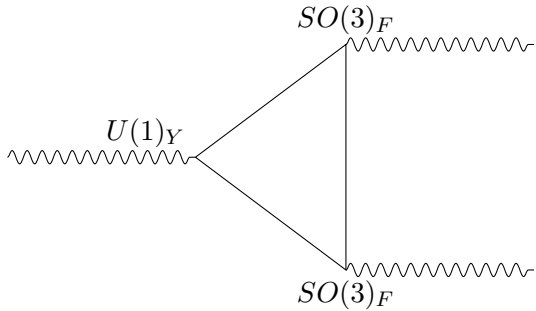

The second is that of anomalies. The most natural way to implement the
model would be to gauge the $SO(3)_F$ symmetry.  This avoids possible
issues with breaking due to gravity, as well as eating any massless
Goldstone bosons.  If $SO(3)_F$ were a global symmetry, there would be
Goldstone bosons that would have to be extremely weakly coupled to the
standard model fields in order to not have been detected.  Even though
they are not directly coupled in this model, it is unclear that, at
loop level, the couplings remain small enough to evade bounds.
If the symmetry were gauged, however, it would induce a $U(1)_Y$
anomaly via the triangle diagram in Figure $\ref{tri-anomaly}$.  Using
the Casimir square operator for the 5 dimensional representation,
$C(5) = 10$, the anomaly is given by
\begin{equation}\label{anomaly-coef}
\mathcal{A}^{ab} = \sum_\ell Y_\ell{\rm Tr}\left(\{t^a_\ell,t^b_\ell\}\right) -
\sum_r Y_r{\rm Tr}\left(\{t^a_r,t^b_r\}\right) = 12 \delta^{ab},
\end{equation}
where $\ell$ are left-handed fermions and $r$ are right-handed
fermions.  Such anomalies can be eliminated by introducing new
fermions. Note the need to introduce new fermions in the full model
once the quarks are included.  The additional fermions may lead to new
light states.  It is beyond the scope of this work to resolve
this issue and to present an anomaly-free model for an $SO(3)$
flavor gauge theory.

Our last remark is about possible variation of our model. Our model
is minimal in many ways, like the choice of the gauge group, the
scalars that breaks $SO(3)_F$, and the fields we choose.  It is likely
that in order to achieve the desired vacuum alignment further
structure would be necessary, including addition symmetries and matter
content.  Furthermore, in our model, the origin of the $Z_2$ symmetry
is unexplained.  However, Abelian discrete symmetries are easier to
produce naturally in the context of orbifolds or spontaneous symmetry
breaking.  Finally, no attempt has been made to incorporate
solutions to the hierarchy problem or other extensions of the Standard
Model.  In particular, the model has not been made supersymmetric and
is four dimensional, while many current models using $A_4$ symmetry
work in supersymmetric theories \cite{XSA4,Altarelli:2009kr} or
theories with extra dimensions \cite{Altarelli:2005yp,Csaki:2008qq}.
It should be possible to extend our model to fit within the structure
of these theories.

\section*{Acknowledgments} 
We thank Csaba Cs\'aki and Yael Shadmi for helpful discussions. This
work is supported by the NSF grant PHY-0757868.

\appendix
\section{Mathematics of $A_4$}\label{app-math}

The non-Abelian discrete group $A_4$ arises in the context of
neutrinos as described in section \ref{a4models}.  For the purposes of
this paper, we would like to be able to determine the irreducible
representations of this group, to determine the result of products of
representations, and to decompose reducible representations into
irreducible representations.  Accomplishing these goals requires
some mathematical background.

The group $A_4$ is defined to be the group of even permutations of $4$
objects.  It is isomorphic to the group of rotational symmetries of
the tetrahedron.  The latter description will be used throughout this
work.  The group is of order $12$ with the elements given as follows:
\begin{itemize}
\item
  The identity 1;

\item
  Rotations by $180^\circ$ about three orthogonal axes (edge-to-edge);

\item
  Rotations by $120^\circ$ and $240^\circ$ about 4 different axes
  (vertex-to-face).
\end{itemize}
This description gives the defining representation, which clearly has
dimension 3 and indicates that $A_4$ is a subgroup of rotations in 3
dimensions $SO(3)$.  Typically, a basis is chosen where the two
generators $S$ and $T$ are given by: 
\beq \label{SandT}
S =
\begin{pmatrix} 
1 & 0 & 0\\ 
0 & -1 & 0\\ 
0 & 0 & -1\\
\end{pmatrix}, \qquad
T = 
\begin{pmatrix}
0 & 1 & 0\\
0 & 0 & 1\\
1 & 0 & 0\\
\end{pmatrix}.
\eeq
This basis is chosen such that the three $180^\circ$ rotation axes are the
Cartesian coordinate axes.

Two irreducible representations are immediately seen at this point:
the defining dimension 3 representation described above and the
trivial representation 1.  There are two more irreducible
representations of $A_4$.  The $1'$ and the $1''$
are dimension 1 representations that map the $120^\circ$ rotations
onto $\omega = e^{2\pi i/3}$ and $\omega^* = e^{4\pi i/3}$
respectively.  The number $\omega$ is a cube root of 1 and satisfies
\begin{equation}
1 + \omega + \omega^2 = 0.
\end{equation}
Notice that these representations are not real. The combination $1'
\oplus 1''$, however, is a real representation isomorphic to the group
generated by a $120^\circ$ rotations in 2 dimensions.  Thus, any real
representation of $A_4$ must contain $1'$ and $1''$ in equal
multiplicities.

The products of these representations are as follows: 
\beqa 
1' \times 1' &=& 1'', \qquad 1' \times 1'' = 1, \qquad 1'' \times 1''
= 1',\qquad 1' \times 3 = 3,\nonumber \\ 1'' \times 3 &=& 3,\qquad 3
\times 3 = 3_1 + 3_2 + 1 + 1' + 1''.  \eeqa Given two triplets
$(x_1,x_2,x_3)$ and $(y_1,y_2,y_3)$, the results of the multiplication
of $3 \times 3$ gives 
\beqa 1 &=& x_1 y_1 + x_2 y_2 + x_3 y_3,
\nonumber \\ 1' &=& x_1 y_1 + \omega x_2 y_2 + \omega^2 x_3 y_3,
\nonumber \\ 1'' &=& x_1 y_1 + \omega^2 x_2 y_2 + \omega x_3 y_3,
\nonumber \\ 3_1 &=& (x_2 y_3, x_3 y_1, x_1 y_2), \nonumber \\ 3_2 &=&
(x_3 y_2, x_1 y_3, x_2 y_1).  
\eeqa 
Furthermore, for a $1'$ (denoted by $u$) and an $1''$ (denoted by $v$),
the multiplications $3 \times 1'$ and $3 \times 1''$ give respectively
\begin{equation}
3 = u (x_1, \omega x_2, \omega^2 x_3),\qquad 3 = v (x_1, \omega^2 x_2,
\omega x_3).
\end{equation}

\begin{table}[!t]
  \begin{center}
    \begin{tabular}{|c|c c c c|}
      \hline
      ~ & $0^\circ$ & $120^\circ$ & $240^\circ$ & $180^\circ$ \\
      \hline
      $\chi_1$ & 1 & 1 & 1 & 1\\
      $\chi_2$ & 1 & $\omega$ & $\omega^2$ & 1\\
      $\chi_3$ & 1 & $\omega^2$ & $\omega$ & 1\\
      $\chi_4$ & 3 & 0 & 0 & $-1$\\
      \hline
    \end{tabular}
    \caption{The character table for $A_4$, listing the conjugacy classes
      on the horizontal and the representations on the vertical.  Here
      $\omega$ satisfies the equation $\omega^2 + \omega + 1 = 0$.  The
      table is taken from \cite{Ledermann:1987}.}\label{char-table}
  \end{center}
\end{table}

Next we need a way to decompose reducible representations of $A_4$
into a direct sum of irreducible representations.  In order to do this
decomposition, we use a theorem about the characters of an element of
a representation.  Given an arbitrary group $G$, an element $g \in G$,
and a representation $\rho$ of $G$, the character is defined as
\begin{equation}
\chi_\rho (g) = {\rm Tr}\rho(g).
\end{equation}
Since the trace is invariant under similarity transformation, every
element of a given conjugacy class will have the same character.
There are four conjugacy classes for $A_4$ given by each of the four
possible angles of rotation: $0^\circ$, $180^\circ$, $120^\circ$, and
$240^\circ$.  The number of conjugacy classes is the same
as the number of irreducible representations.  This is a general
result that holds for any finite group.  It allows the
construction of a character table listing the characters by
irreducible representation and conjugacy class.  For $A_4$, the
character table is given in Table \ref{char-table}.  

Given a representation $\rho$ which is not necessarily irreducible,
irreducible representations $\rho_i$ and an element $g \in G$, the
following relation holds:
\begin{equation}\label{irrep-conj}
\chi_\rho(g) = \sum_i n_i \chi_{\rho_i} (g),
\end{equation}
where $n_i$ is the multiplicity of $\rho_i$ in the decomposition of
$\rho$ into irreducible representations. In the case of $A_4$,
$i=1,1',1'',3$. Notice that the number of multiplicities $n_i$ is
given by the number of irreducible representations of $G$.  Such an
equation can be written down for each conjugacy class of $G$.  Thus,
if we wish to determine the multiplicities $n_i$, we have the same
number of variables as equations given by (\ref{irrep-conj}).  Given
the characters in the representation under study and the irreducible
representations, it is then possible to determine the decomposition of
the representation $\rho$ into irreducible representations.  The
characters of the irreducible representations are given by the
character table.  The characters of the representation under study can
be computed directly.  In our case, we are interested in studying the
representations of $A_4$ induced by irreducible representations of
$SO(3)$.  In this case, computing the characters is even simpler as a
general formula for the characters in $SO(3)$ has been
determined~\cite{Etesi:1997jv}:
\begin{equation}\label{char-so3}
\chi_j (\theta) =
\frac{\sin\left[(2j+1)\theta/2\right]}{\sin\left(\theta/2\right)},
\end{equation}
where $j$ is the spin of the representation and $\theta$ is the angle
of rotation.

\begin{table}[!t]
  \begin{center}
    \begin{tabular}{|c|c c c c|}
      \hline ~$j$~ & ~$n_1$~ & ~$n_{1'}$~ &
      ~$n_{1''}$~ & ~$n_3$~\\ 
      \hline 
      $0$ & $1$ & $0$ & $0$ & $0$\\
      $1$ & $0$ & $0$ & $0$ & $1$\\
      $2$ & $0$ & $1$ & $1$ & $1$\\
      $3$ & $1$ & $0$ & $0$ & $2$\\
      $4$ & $1$ & $1$ & $1$ & $2$\\
      $5$ & $0$ & $1$ & $1$ & $3$\\
      \hline
    \end{tabular}
    \caption{Decomposition of the six smallest representations of
      $SO(3)$ into irreducible representations of $A_4$.  The 4
      rightmost columns indicate the multiplicity of the four
      irreducible representations of $A_4$.}\label{decomp-6}
  \end{center}
\end{table}

For a spin $j$ representation of $SO(3)$, the decomposition under
$A_4$ proceeds as follows.  There are four conjugacy classes of $A_4$,
corresponding to rotations by $0^\circ$, $180^\circ$, $120^\circ$, and
$240^\circ$.  The characters of these rotations under the
representation of $SO(3)$ are given by (\ref{char-so3}).  The
multiplicities of $1'$ and $1''$ must be equal
since the group $SO(3)$ is real.  Then, using (\ref{irrep-conj}), the
following set of equations can be written:
\beqa
  2j + 1 &=& n_1 + 2 n_{1'} + 3 n_3, \nonumber \\
  (-1)^j &=& n_1 + 2 n_{1'} - n_3, \nonumber \\
  \frac{2}{\sqrt{3}}\sin\frac{(2 j + 1)\pi}{3} &=& n_1 - n_{1'} +
  \omega^2 n_3 
\eeqa
Note that the last two equations are cyclic in $j$ with period 6.
This results in a pattern with that period.  The decomposition for the
first six representations is given in Table~\ref{decomp-6}.  The
pattern for a higher representation $j$ can be determined as follows.  Let
\beq
q = \lfloor j/6 \rfloor, \qquad r = j~{\rm mod}~6.  
\eeq
Then for $i=1,1',1''$ we have
\beq
n_i(j)=n_i(r) + q. 
\eeq
For $i=3$ we have
\beq
n_3(j)=n_3(r) + 3 q. 
\eeq
For example, the spin $j=23$ representation has $q = 3$ and $r = 5$,
and thus $n_1(23)=3$, $n_{1'}(23)=n_{1''}(23)=4$ and $n_3(23)=12$.

\section{Minima of the potential of a $\mathbf{7}$ of $SO(3)$}\label{app-min}

In this appendix, we present the determination of the minima of the
potential (\ref{pot-7}) as done in \cite{Ovrut:1977cn}.  In order to
proceed, it is simplest to reparametrize $T^{abc}$ based on
symmetries. Before we do that, however, we start with a simpler
example: the case of a triplet.  We can write the $\mathbf{3}$ as the
product of a magnitude and a unit vector: $v^a = \alpha x^a$ such that
the three parameters are the length of $v$, denote by $\alpha$, and
the two angles that describe the orientation of $v^a$.  The point to
emphasize is that the potential for such a scalar is written as a
function of only one of the parameters, the magnitude $\alpha$. It is
given by
\begin{equation}\label{pot-3}
V = -\frac{\mu^2}{2} v^a v^a + \frac{\lambda}{4!} (v^a v^a)^2 =
-\frac{\mu^2}{2} \alpha^2 + \frac{\lambda}{4!} \alpha^4.
\end{equation} 
Furthermore, if $\mu^2>0$ and $\lambda>0$, the resulting vacuum has the
residual symmetry of the unit vector $x^a$, which is $SO(2)$.

For the $\mathbf{7}$, the parametrization and potential are both more
complicated.  There are three orthogonal terms with different
symmetries.  The first term is invariant under $SO(2)$ as it depends
on a single unit vector.  The second term is best described
geometrically.  Consider an arbitrary equilateral triangle in three
dimensions.  Define three vectors connecting the center of the
triangle to each of the three vertices of the triangle.  The object
defined by these vectors is called a regular 3-point star.
Mathematically, it can be written as the symmetric outer product of
the three defining vectors.  This construction is automatically
traceless.  The second term is then given by a regular 3-point star
defined with unit vectors.  Finally, the third term is given by the
symmetric product of three orthonormal unit vectors.

Explicitly, the parametrization is
\begin{equation}\label{param}
T^{abc} = \alpha \left(x^a x^b x^c - \frac{3}{5} \delta^{(ab}
x^{c)}\right) + \beta \chi^{abc}_{(3)} + \gamma x^{(a} y^b z^{c)},
\end{equation}
where $\chi^{abc}_{(3)}$ describes an arbitrary 3 point regular star
with unit length vectors, the vectors $x$, $y$, $z$ are orthonormal
and $\chi$ is orthogonal to $x$.  A general tensor written as in
(\ref{param}) has 7 parameters as one would expect for a symmetric
traceless tensor of rank 3: $\alpha$, $\beta$, $\gamma$, the two
angles in $x^a$, the angle of $\chi_{(3)}$ about the $x$ axis, and the
angle of $y$ about the $x$ axis.  The angle of $z$ is determined by
requiring orthogonality.  There are two advantages to the
parametrization (\ref{param}). The first is that since the terms are
orthogonal and normalized, the potential can now be written in terms
of the three parameters $\alpha$, $\beta$, and $\gamma$ rather than in
terms of seven parameters.  The second is that, once the vacua are
determined, it is far easier to determine the symmetries in this
parametrization.  The three terms in the parametrization have
well-defined symmetry groups.  The first is invariant under $SO(2)$
(rotations orthogonal to $x$).  The second is invariant under $D_3$
since a three point star has the symmetries of a triangle.  The third
is invariant under $A_4$, where the three vectors $x$, $y$, and $z$
are taken to be the $180^\circ$ rotation axes.  If the basis is chosen
such that $x$, $y$, and $z$ are along the corresponding axes of the
coordinate system, this term is invariant under both $S$ and $T$ given
in (\ref{SandT}).

The potential of Eq. (\ref{pot-7}) written in terms of (\ref{param})
depend only of three out of the seven parameters, $\alpha$, $\beta$,
and $\gamma$. It is given by
\begin{multline}
  V = - \frac{\mu^2}{2} \left(\frac{2}{5} \alpha^2 + \frac{1}{4} \beta^2 +
  \frac{1}{6} \gamma^2\right) + \frac{\lambda}{4} \left(\frac{2}{5}
  \alpha^2 + \frac{1}{4} \beta^2 + \frac{1}{6} \gamma^2\right)^2 +\\ c
  \left(\frac{44}{25^2} \alpha^4 + \frac{1}{25} \alpha^2 \beta^2 +
  \frac{2}{25} \alpha^2 \gamma^2 + \frac{1}{24} \beta^2 \gamma^2 +
  \frac{3}{18^2} \gamma^4\right).
\end{multline}
In order for an $A_4$-invariant vacuum to exist, there must be a
minimum with $\alpha = \beta = 0$ and $\gamma \neq 0$.  Indeed, there
is such a minimum for a certain portion of parameter space.  If $c >
0$, then there is a $D_3$ invariant vacuum (only $\beta \neq 0$).  For
$- \lambda/2 < c < 0$, there is an $A_4$ invariant vacuum.  Finally, for
$c < - \lambda/2$, the potential has a runaway direction.  It is
possible to spontaneously break $SO(3)$ to $A_4$ using a single scalar
in a spin 3 representation of $SO(3)$ by picking the second case,
$- \lambda/2 < c < 0$.


\end{document}